 \documentclass[12pt,preprint]{aastex}

\slugcomment{To appear in Astrophysical Journal}

\shorttitle{Variability in Apparently Single Wolf-Rayet Winds I.}
\shortauthors{St-Louis et al.}

\begin{document}


\title{A Systematic Search for Corotating Interaction Regions in Apparently Single Galactic WR Stars:\\
I. Characterizing the Variability.}


\author{N. St-Louis}
\affil{D\'{e}partement de Physique, Universit\'{e} de Montr\'{e}al,
C.P. 6128, Succ. Centre-Ville, Montr\'{e}al, Qu\'{e}bec, Canada H3C 3J7}
\email{stlouis@astro.umontreal.ca}

\author{A.-N. Chen\'e}
\affil{Canadian Gemini Office, Herzberg Institute of Astrophysics, \\ 5071, West Saanich Road,  Victoria (BC), Canada V9E 2E7}
\email{andre-nicolas.chene@nrc-cnrc.gc.ca}

\author{O. Schnurr}
\affil{Department of Physics \& Astronomy, University of Sheffield, Hicks Building, Hounsfield Rd, Sheffield, S3 7RH, UK}
\email{o.schnurr@sheffield.ac.uk}

\and

\author{M.-H. Nicol}
\affil{Max Planck Institute for Astronomy, Konigstuhl 17, D-69117, Heidelberg, Germany}
\email{nicol@mpia.de}



\begin{abstract}
We present the results of a systematic search for large-scale
spectroscopic variability in apparently single Wolf-Rayet stars
brighter than $v\sim12.5$. In this first paper we characterize the various 
forms of variability detected and distinguish several separate groups.
For each star in our sample, we obtained 4--5
high-resolution spectra with a signal-to-noise ratio $\sim$~100. Our
ultimate goal is to identify new candidates presenting variability that potentially
comes from Co-rotating Interaction Regions (CIR).

Out of a sample of 25 stars, 10 were found to display large-scale changes
of which 4 are of CIR-type (WR\,1, WR\,115 WR\,120 and WR\,134).  The star WR\,134 
was already known to show such changes from previous studies. Three WN8 stars 
present a different type of large-scale variability and we believe
deserve a group of their own. Also, all three WC9d stars in our
sample present large-scale variability, but it remains to be checked if
these are binaries, as many dust-making WR stars are double. Finally, of the
remaining stars, 10 were found to show small-amplitude spectral changes which we
attribute to normal line-profile variability due to inhomogeneities in the wind, 
and 5 were found to show no spectral variability, as far as can be concluded 
from the data in hand.

Follow-up studies are required to identify potential periods for our candidates 
showing CIR-type changes and eventually estimate a rotation rate for these WR stars.
 
\end{abstract}





\section{Rotation and Spectroscopic Variability of Single Wolf-Rayet Stars}

Stellar rotation has now been fully incorporated into evolutionary
models of massive stars \citep[eg.][]{Mey00,Heg00},  prompted
by several observational facts that were unexplained by previous
models without rotation \citep[eg.][]{Mey94,Wosetal93} and that
pointed to a solution related to mixing. For example, \citet{Her92}
had found that many O stars, and in particular fast rotators, show an
enhanced He abundance; they called this the {\it helium
discrepancy}. The models including rotation have been very successful
in reproducing the observed abundances, predicting higher He and N
abundances for higher masses and higher rotation
velocities. Evolutionary models with rotation have also provided a
solution for many other previously unexplained observations such as
the well-known blue-to-red supergiant ratio (B/R) problem
\citep[e.g.][]{Lan95}. Indeed, when neglecting rotation, evolutionary
models were unable to reproduce the rapid increase in B/R with
increasing metallicity.  As the new models predict many more red
supergiants at low metallicity coupled with a shorter blue supergiant
phase, they provided a natural solution.

It is now clear that rotation is an important factor in our
understanding of massive stars and their evolution.  It is therefore
interesting to compare theoretical predictions with the available
observational data.  For O stars, the latest homogenous, large-scale
surveys of projected rotational velocities have been carried out by
\citet{Pen96} and \citet{How97}, who respectively studied 177 and 373
stars, the latter also including early-type B stars. These studies
were both based on high-resolution IUE spectra and used a
cross-correlation technique with a narrow-line template spectrum to
measure the rotational speeds. Both studies note the absence of very
narrow lines for supergiants while some smaller velocities are found
for main-sequence stars. Instead of interpreting this as a possible
increase in rotational velocities as the star evolves, these authors
favour an additional broadening mechanism such as
macro-turbulence. In fact, \citet{How04} strengthens this assertion by
pointing out that the shape of the absorption lines expected from pure
rotation is a poor match to the observed profiles and that a line
dominated by transverse turbulence is a much better match. Ignoring 
this possible effect (most likely small) of macro-turbulence on the projected
rotational velocities, their distribution peaks at $\sim$ 80
km\,s$^{-1}$ and has a very extended high-velocity tail reaching to
400 km\,s$^{-1}$.  These values are not incompatible with model
predictions; on the main sequence, the models show that even if the
star starts out with a relatively flat internal rotation profile, the
rotation becomes increasingly differential (but never more than a
factor of two in velocity) because the convective core is contracting
and therefore spinning up, while the outer layers are expanding and therefore
slowing down. In spite of this, there is a global decrease in the
rotation rate as a function of time as angular momentum is lost
through mass-loss. So, depending on their age and their initial
rotation rate, the observed velocities are quite
acceptable. Interestingly, \citet{Pen96} notes that, in fact, the
higher rotational velocities are generally associated with lower-mass
stars. This is an important result. Indeed \citet{Mey00} predict
that the rotation rates decrease more rapidly for higher initial rotation
rates and for higher mass and therefore higher mass-loss rate stars, in
view of the increased loss of angular momentum. This agrees very
nicely with the observations.

For WR stars, very slow average surface rotation velocities in the range
20--70 km\,s$^{-1}$ are predicted for the He-burning phase, depending on
the initial rotation velocity on the main sequence and on the evolutionary
state of the star \citep{Mey03}.  The reason for these low values and the narrowness 
of this range is simple; the amount of mass that must be removed
from massive O stars to turn them into WR stars is so big that most of
the angular momentum is lost in the process.  However, the
differential rotation that steadily increased on the main sequence
suddenly becomes very pronounced, as the helium core contracts at the
end of core hydrogen burning. The star ends up with a core spinning
faster than the initial ZAMS rotation velocity.

A fast-spinning core, at least in some WR stars, would support the
presently favoured model to explain long-soft Gamma-Ray Bursts
(GRBs). Indeed, the so-called {\it collapsar} model
\citep{Wos93,Mac99,Mac01} involves the collapse of a rapidly rotating
massive star in its pre-supernova phase. In this model, the central
core becomes a black hole surrounded by a disk. The accretion of
matter through the disk creates powerful jets in which the matter
travels at very high speeds. \citet{HegLan00} and \citet{Hir04}, have
followed the evolution of a rotating He core to determine if the very
rapid core velocities required by the collapsar model could be
reached in spite of the many modes of transport and redistribution of
the angular momentum taking place (mass-loss, shear, turbulence,
convection, etc). They found that indeed this is possible for initial
rotation speeds on the main sequence that are typical of that of O
stars.

However, adding magnetic fields to the models has brought new elements 
to the problem. Indeed, \citet{Heg05} have shown that including magnetic 
torques caused by a Spruit-Tayler dynamo \citep{Spr02} dramaticaly 
reduces the angular momentum of the core leading to periods in the resulting 
pulsar of 15 ms at birth, which is compatible with the average value for common 
pulsars but much too small to make a long-soft gamma-ray burst (GRB).

Nevertheless, for a rotating star, efficient chemical mixing can occur and if the
mixing timescale is shorter than the thermonuclear timescale, the star can remain 
quasi-chemically homogeneous \citep{Mae87,Lan92}.  In that case, the star lacks 
the stratified layers which are required for the development of the Spruit-Tayler 
dynamo \citep{Spr02}.  This lack of a core-envelope structure prevents the magnetic 
core-envelope coupling described above and the spin-down of the star's core can be 
avoided. \citet{Wos06} have shown that for a rapidly-rotating star on the main sequence and if 
the mass-loss in the WR phase is reduced by a factor of ten, massive-star evolution 
can indeed lead to a core with sufficient angular momentum to produce a long-soft 
GRB.  The necessary reduced mass-loss in the WR phase favors 
low metallicity environments for GRBs. Finally \citet{Yoo06} calculated evolutionary 
grids for massive stars including rotation and the effects of a Spruit-Tayler dynamo for different metalicities.  
Still adopting WR mass-loss rates reduced by a factor of ten, they predict that the 
production of long GRBs should be limited to metalicities smaller than $Z\sim$ 0.004.  
This translates into a peak at a redshift of z=4.

No reliable direct observation of WR rotation rates has ever been
obtained. Attempts have been made to estimate them based on the width
of (what the authors had assumed to be) photospheric absorption lines in WR spectra
\citep{Mas80,Mas81}, but these have been seriously questioned, as wind expansion and
turbulence have not been taken into account. 

It may be virtually impossible to obtain the rotationnal velocity of WR stars
directly through photospheric line-profiles but it could still be possible to deduce it 
from periodic variability in photometry or of spectral lines formed in the winds. 
Indeed, the winds of massive stars have been demonstrated to be highly
variable.  In particular, one type of structure in the wind that
accounts for some of these variations, Co-rotating Interaction Regions
(CIRs), are thought to be closely linked to the rotation of the
star. Indeed, these structures are caused by perturbations at the
base of the wind that propagate through it and are carried by
rotation.  This generates a spiral-like structure in the density
distribution, analogous to those commonly observed in the solar corona
\citep{Hun72}, which very likely leads to the ubiquitous Discrete
Absorption Components in the spectra of O stars \citep{Cran96}. These
are rarely observed in WR P~Cygni profiles as their absorption
components are usually saturated \citep[one exception is
WR\,24;][]{Pri92}.  However, it is believed that they also lead to a
very characteristic, large-scale, periodic variability pattern in
WR-wind emission lines. \citet{Des02} show the unambiguous S-shape 
pattern in time-series of model spectra of WR emission lines produced 
as a consequence of these variations. Observational
support has been obtained for this type of variability in two clear
cases through repeated spectroscopic observations.  WR\,6
\citep[][P=3.76 days]{StL95,Mor97} and WR\,134 \citep[][P=2.25 days]{McC94,Mor99}
show periodic variations without any indication of a companion. 
The correspondence between the data and models
is striking when one considers that optical depth effects, which are
certainly occurring in the observed data, have not been included in
the models. This brings strong support to the idea that CIRs do in
fact trace the {\bf rotation rate} of the star at the base of the wind.
According to \citet{Ham88,Ham95}, these stars both have 
$R_*  \simeq 3 R_{\odot}$.  This radius corresponds by definition
to a Rosseland optical depth of 20.  Assuming that CIRs form at this
radius, the equatorial rotation speed (at $R_*$) is $\sim$ 40 km s$^{-1}$ 
for WR\,6 and $\sim$ 70 km s$^{-1}$ for WR\,134, not too different from 
values predicted by evolutionary models. The periods and variability
patterns were found to be the same for lines formed at various
distances in the wind, indicating that the CIRs are governed by angular 
momentum conservation which results in a coherent structure in the wind.
The period therefore provides a direct measurement of the rotation 
{\em rate} of the underlying star. To calculate the rotation velocity we need to know 
at what radius the perturbation leading to the CIR originates.  If this radius
is known, we can then calculate the rotation velocity {\em at that radius}.
Much more observations and analysis of variability associated with CIRs are 
required to reach this goal. Therefore, the above velocities are only given 
as rough estimates for illustrative purposes.

In an attempt to identify more WR stars displaying CIR-type
variability, we have carried out a survey of all Galactic WR stars
brighter than $v\sim12.5$.  Our complete sample includes 39 southern
stars, which will be discussed in \citet{Che09a}, and 25 northern stars,
which we discuss here.  The goal of this paper is to present the
different types of spectroscopic variability found in our magnitude-limited
sample of stars in a qualitative but also quantitative manner.  Our aim
is to use these characteristics to establish some guidelines to help identify 
the spectroscopic variations associated with CIRs.  Once identified, these stars
can then be observed more intensively and, if present, periods can be determined.  
In our second paper, we will build on these basis and use the complete sample to
determine the frequency of the various types of line-profile variability.  Follow-up
papers on individual stars that have been more intensively monitored are also planned. 

Our northern sample includes all Galactic WR
stars visible from the Mont M\'egantic Observatory that are not
confirmed as SB1 or SB2 binaries. Our targets are listed in Table~1 where
the stars' name, spectral type, RA, DEC and $v$ are provided. When available, we also
provide terminal velocities measured either from IUE absorption troughs \citep{Pri90},
from near-infrared line profiles \citep{How92} or from optical line profiles \citep{Ham06}. The
last column indicates each star's variability status (see below). Note
that although some stars in our sample have published spectral types that indicate a possible
companion (not indicated in the table), none have confirmed orbits. In
the case of WR\,4 and WR\,128, only photometric variations have been
found without radial velocity confirmation of binarity.
In the case of WR\,131, WR\,143, WR\,156 and
WR\,158, the claim of binarity is based solely on diluted emission
lines (d.e.l.) and/or the presence of absorption lines in the spectrum
(+a). However, \citet{Ham07} argue that binary status established from
the d.e.l or +a criteria is often misleading as WR stars can produce a
wide variety of spectra depending on their physical parameters. Those
authors have demonstrated that in many cases these types of spectra
can be reproduced entirely with a single-star model. In particular,
WR\,108, WR\,156 and WR\,158 were among the stars that they reproduced
with a single-star model.  Our final sample consists of 16 WN stars
and 9 WC stars. We have included WR\,134 for comparison even though it
is already known to show variability thought to be associated with CIRs.
However, we have excluded WR\,157 because it is too close on the sky to another
star that is only $\sim$1 magnitude fainter.  The separation between the two stars is
$\sim$2.2$''$, i.e. a value comparable with the average seeing at the mont M\'egantic
Observatory, where our data were obtained.  It was therefore impossible to isolate the WR star
in our 1.5$''$ spectrograph slit.

\begin{deluxetable}{ccccccccc}
\tabletypesize{\scriptsize}
\tablecolumns{9}
\tablewidth{0pc}
\tablecaption{Our Sample of Single WR Stars in the Northern Hemisphere}
\tablehead{
\colhead{Name} & \colhead{Spectral Type\tablenotemark{1}} & \colhead{RA (2000)} & \colhead{DEC (2000)} & \colhead{$v$} & \multicolumn{3}{c}{$v_{ \infty}$(km\,s$^{-1}$)}& \colhead{Variability}  }
\startdata
WR1&WN4&00 43 28.4&+64 45 35.4&10.51&2100\tablenotemark{3}&1900\tablenotemark{4}&&LSV\\
WR2&WN2&01 05 23.03&+60 25 18.9&11.33&3200\tablenotemark{3}&1800\tablenotemark{4}&&NV\\
WR3&WN3&01 38 55.63&+58 09 22.7&10.70&&2700\tablenotemark{4}&&SSV\\
WR4&WC5&02 41 11.68&+56 43 49.7&10.53&&&&NV\\ 
WR5&WC6&02 52 11.66&+56 56 07.1&11.02&&&&NV\\
WR106&WC9d&18 04 43.66&-21 09 30.7&12.33&&&&LSV\\
WR108&WN9h&18 05 25.74&-23 00 20.3&10.16&1220\tablenotemark{3}&1170\tablenotemark{4}&&SSV\\
WR110&WN5-6&18 07 56.96&-19 23 56.8&10.30&3190\tablenotemark{3}&2300\tablenotemark{4}&&SSV\\
WR111&WC5&18 08 28.47&-21 15 11.2&8.23&2415\tablenotemark{2}&2300\tablenotemark{3}&&NV\\
WR115&WN6&18 25 30.01&-14 38 40.9&12.32&1480\tablenotemark{3}&1280\tablenotemark{4}&&LSV\\
WR119&WC9d&18 39 17.91&-10 05 31.1&12.41&&&&LSV\\
WR120&WN7&18 41 00.88&-04 26 14.3&12.30&1225\tablenotemark{3}&1225\tablenotemark{4}&&LSV\\
WR121&WC9d&18 44 13.15&-03 47 57.8&12.41&1300\tablenotemark{3}&&&LSV\\
WR123&WN8&19 03 59.02&-04 19 01.9&11.26&970\tablenotemark{4}&&&LSV\\
WR124&WN8h&19 11 30.88&+16 51 38.2&11.58&710\tablenotemark{4}&&&LSV\\
WR128&WN4(h)&19 48 32.20&+18 12 03.7&10.54&2270\tablenotemark{2}&2050\tablenotemark{4}&&SSV\\
WR131&WN7h&20 00 19.12&+33 15 51.1&12.36&1400\tablenotemark{4}&&&SSV\\
WR134&WN6&20 10 14.20&+36 10 35.1&8.23&1905\tablenotemark{2}&1960\tablenotemark{3}&1700\tablenotemark{4}&LSV\\
WR135&WC8&20 11 53.53&+36 11 50.6&8.36&1405\tablenotemark{2}&1500\tablenotemark{3}&&SSV\\
WR136&WN6(h)&20 12 06.55&+38 21 17.8&7.65&1605\tablenotemark{2}&1760\tablenotemark{3}&1600\tablenotemark{4}&SSV\\
WR143&WC4&20 28 22.68&+38 37 18.9&11.95&&&&SSV\\
WR152&WN3(h)&22 16 24.05&+55 37 37.2&11.67&1800\tablenotemark{3}&2000\tablenotemark{4}&&SSV\\
WR154&WC6&22 27 17.82&+56 15 11.8&11.54&2700\tablenotemark{3}&&&NV\\
WR156&WN8h&23 00 10.13&+60 55 38.4&11.09&660\tablenotemark{4}&&&LSV\\
WR158&WN7h&23 43 30.60&+61 55 48.1&11.46&900\tablenotemark{4}&&&SSV\\
\enddata
\tablenotetext{1}{Spectral types are from \citet{Huc}}
\tablenotetext{2}{\citet{Pri90}}
\tablenotetext{4}{\citet{How92}}
\tablenotetext{4}{\citet{Ham06}}
\end{deluxetable}

\section{Observations}

Our observations were obtained at the 1.6m telescope of the mont
M\'egantic Observatory in Qu\'ebec, Canada during 6 runs in 2001 and
2002. Details are given in Table~2 in which the run number,
the dates of observations and the Julian Dates (JD) are listed. We
used a 830.8 l/mm grating giving a resolution of $\Delta\lambda$=1.6 \AA\
(3 pix.) combined with a 2048$\times$4096 EEV CCD.

\begin{table}[h]
\caption{Observing Runs}
\begin{center}
\begin{tabular}{ccc}
\hline
Observing & Dates & Julian Date\\
run No.   &       &            \\
\hline
1&25/06/2001--05/07/2001&2452086--2452096\\
2&26/07/2001--02/08/2001&2452117--2452124\\
3&01/10/2001--07/10/2001&2452184--2452190\\
4&19/06/2002--24/06/2001&2452445--2452450\\
5&17/07/2002--26/07/2002&2452473--2452482\\
6&16/10/2002--24/10/2002&2452564--2452572\\
\hline
\end{tabular}
\end{center}
\end{table}

Our goal is to identify new large-scale variability within our sample
of stars. The extra features we are looking for on top of the main
emission lines are quite large and therefore easy to find.  Based on
our previous experience with WR\,6 and WR\,134, we have made the assumption that
a small number of good quality spectra, well sampled in time are
sufficient to identify new candidates showing variability possibly 
associated with CIRs.  Of course, to quantify
the changes and determine the period, a more thorough follow-up will
subsequently be required.  Therefore, we set out to obtain 4--5 spectra per
star in our sample.

The spectra were reduced using the IRAF software package in the
standard way.  First a bias was subtracted, then each image was
divided by a flat field. After extracting the spectra, the wavelength
calibration was carried out using a He-Ar calibration lamp. Although
our original spectral range was 1500 \AA\  wide, we were forced to reject the
reddest part of the spectrum due to a lack of usable lines in the
calibration spectra. The final wavelength interval is
$\Delta\lambda$=4500-5200 \AA. The signal-to-noise ratio of the spectra
ranges from 100 to 120.

As our spectra are not photometrically calibrated, the final step of
our data reduction procedure was to rectify the spectra.  To do so, we
have fitted a low order Legendre polynomial to wavelength regions
with no strong spectral lines and divided our spectra by the fitted
curves.

\section{Variability Search}

\begin{figure}[htbp]
  \centerline{\plotone{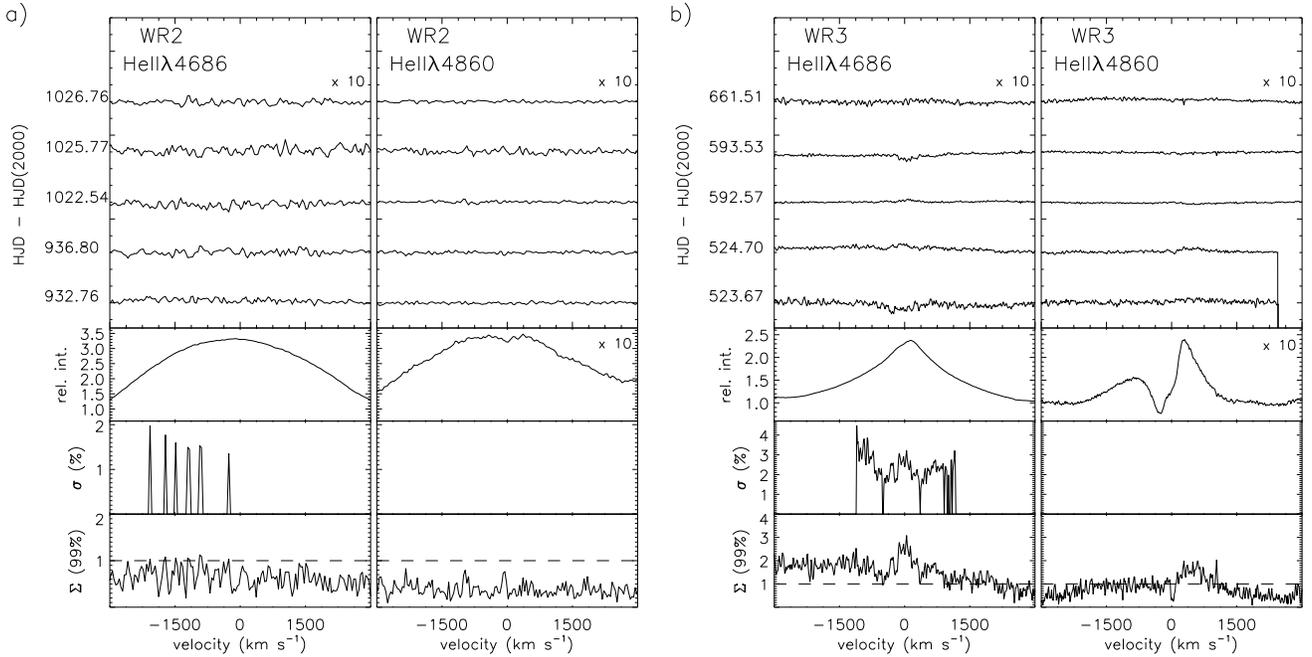}}
  \caption[Montage of line-profiles from a) WR~2 (WN2) and b) WR~3 (WN3ha)]{a) {\it top}: Montage of the He{\sc ii}$\lambda$4686 (left) and He{\sc i}$\lambda$4860 (right) residuals (individual spectra $-$ mean) for WR~2 (WN2). For both cases, the scale factor of the ordinate is indicated in the top right-hand corner of the plot. HJD - HJD(2000) is indicated in the {\it y}-axis. {\it second from top}: Mean spectrum. {\it second from bottom}: $\sigma$-spectrum. {\it bottom}: $\Sigma$~(99\%) spectrum. b) The same as a) for WR~3 (WN3ha).}
  \label{fig1N}
\end{figure}

\begin{figure}[htbp]
  \centerline{\plotone{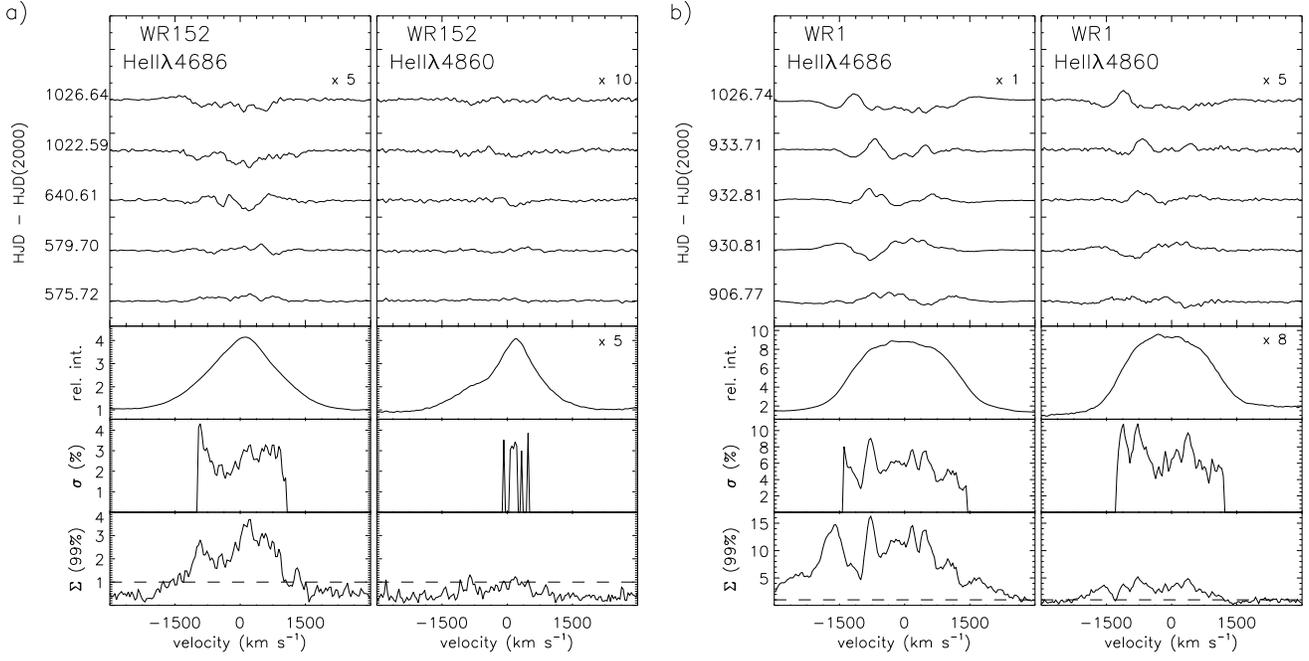}}
  \caption[Montage of line-profiles from a) WR~152 (WN3) and b) WR~1 (WN4)]{Same as Fig.\ref{fig1N} for a) WR~152 (WN3) and b) WR~1 (WN4).}
  \label{fig2N}
\end{figure}

\begin{figure}[htbp]
  \centerline{\plotone{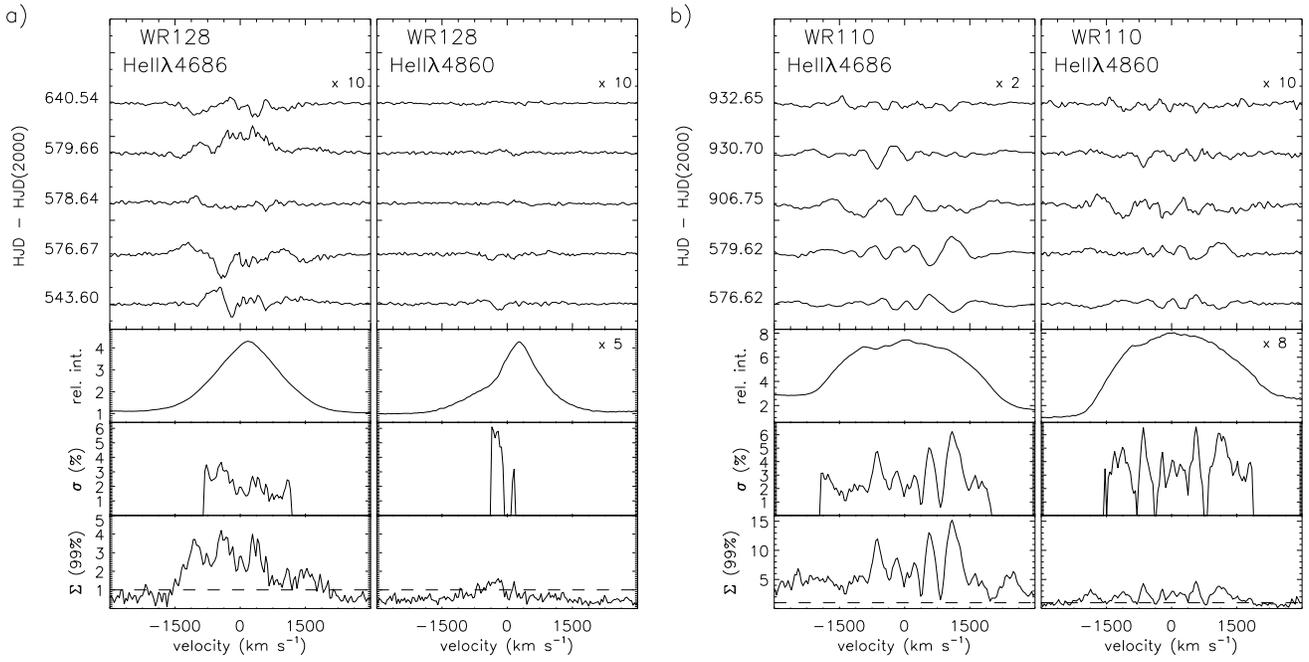}}
  \caption[Montage of line-profiles from a) WR~128 (WN4(h)) and b) WR~110 (WN5--6)]{Same as Fig.\ref{fig1N} for a) WR~128 (WN4(h)) and b) WR~110 (WN5--6).}
  \label{fig3N}
\end{figure}

\begin{figure}[htbp]
  \centerline{\plotone{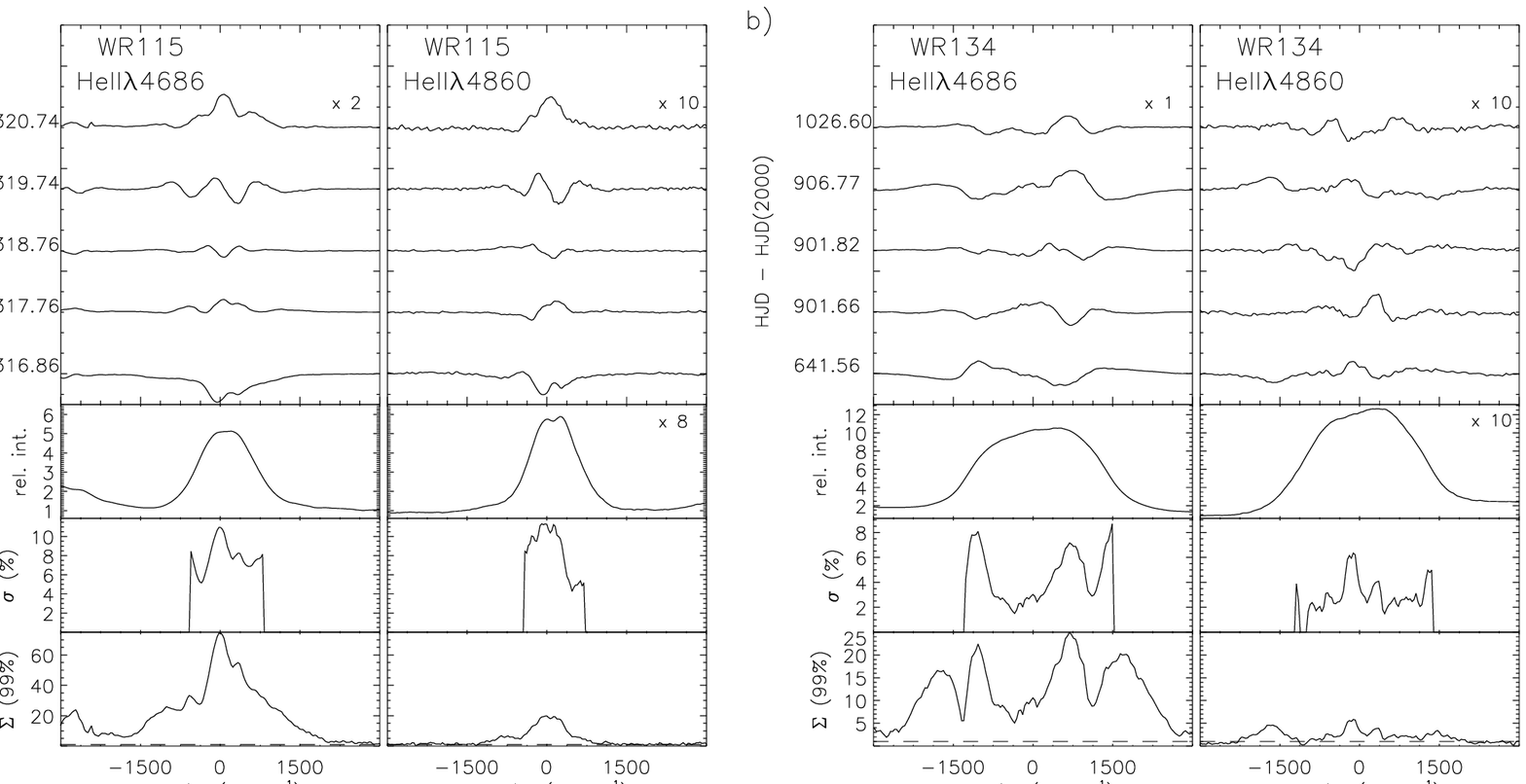}}
  \caption[Montage of line-profiles from a) WR~115 (WN6) and b) WR~134 (WN6)]{Same as Fig.\ref{fig1N} for a) WR~115 (WN6) and b) WR~134 (WN6).}
  \label{fig4N}
\end{figure}

\begin{figure}[htbp]
  \centerline{\plotone{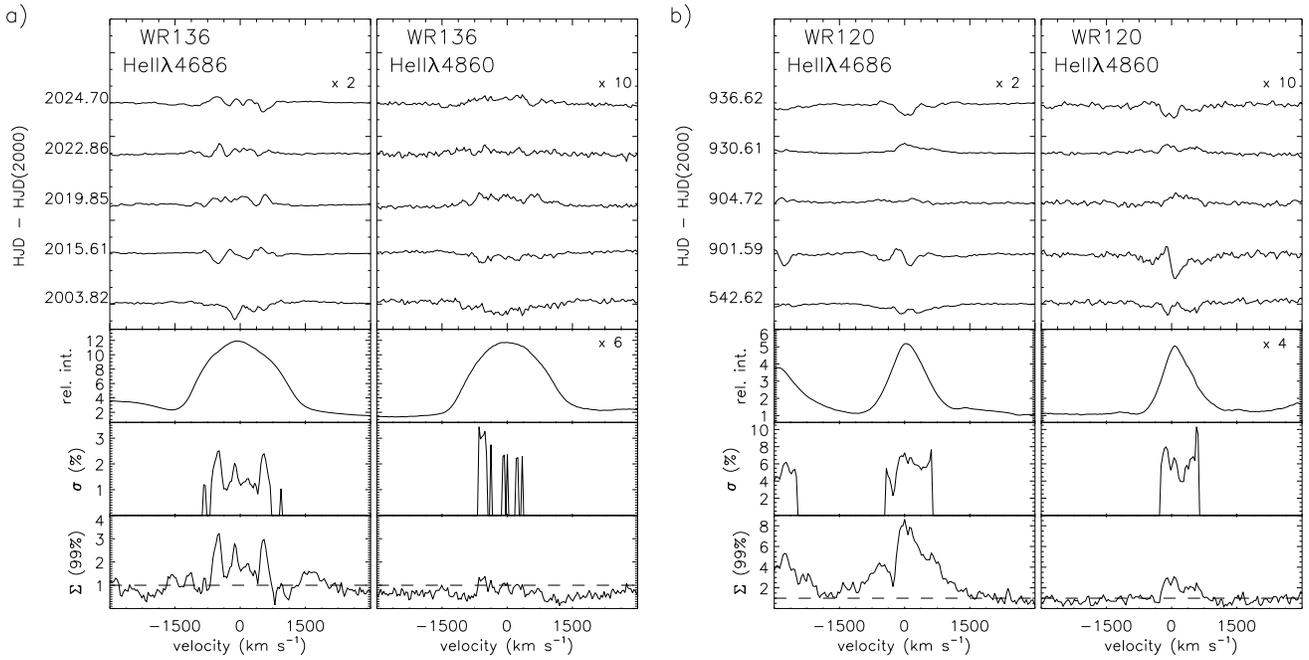}}
  \caption[Montage of line-profiles from a) WR~136 (WN6) and b) WR~120 (WN7)]{Same as Fig.\ref{fig1N} for a) WR~136 (WN6) and b) WR~120 (WN7).}
  \label{fig5N}
\end{figure}

\begin{figure}[htbp]
  \centerline{\plotone{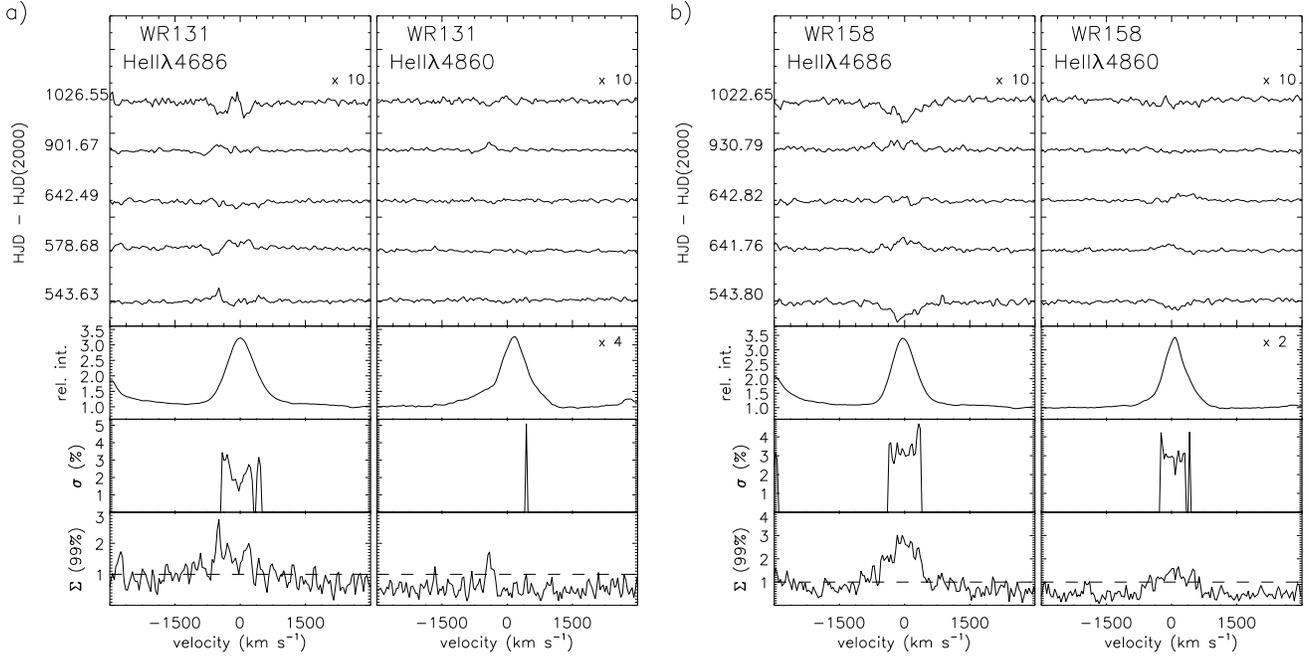}}
  \caption[Montage of line-profiles from a) WR~131 (WN7h) and b) WR~158 (WN7h)]{Same as Fig.\ref{fig1N} for a) WR~131 (WN7h) and WR~158 (WN7h).}
  \label{fig6N}
\end{figure}

\begin{figure}[htbp]
  \centerline{\plotone{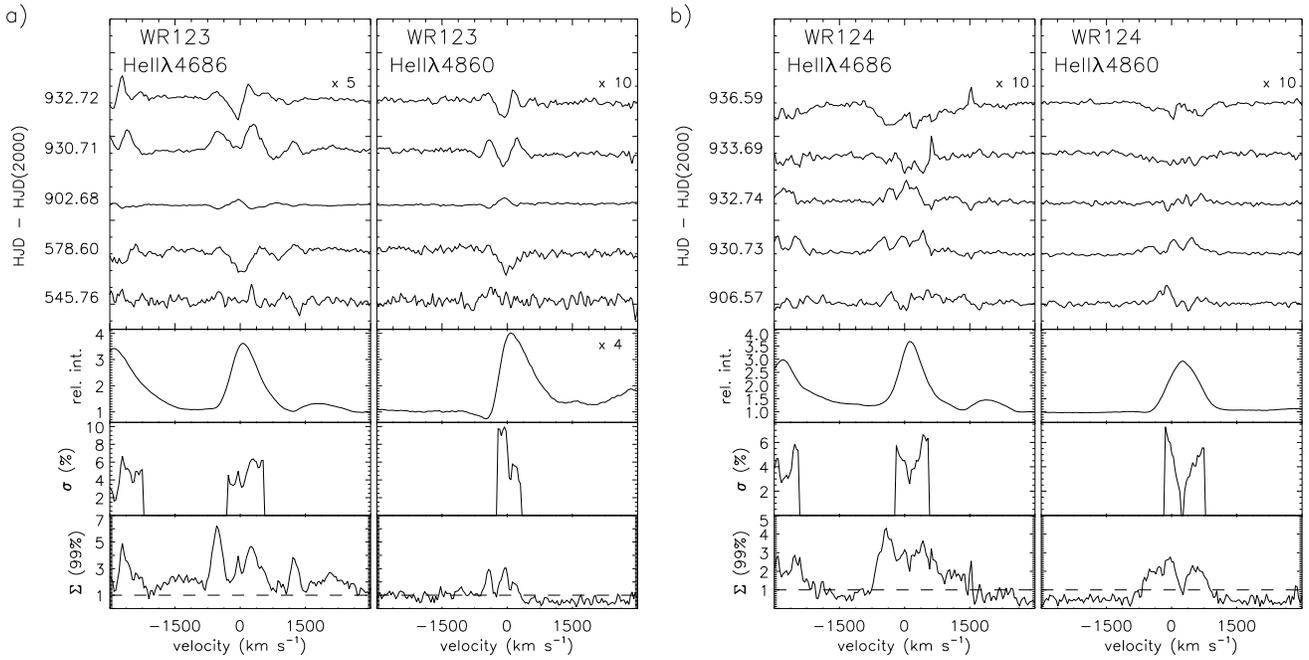}}
  \caption[Montage of line-profiles from a) WR~123 (WN8) and b) WR~124 (WN8h)]{Same as Fig.\ref{fig1N} for a) WR~123 (WN8) and b) WR~124 (WN8h).}
  \label{fig7N}
\end{figure}

\begin{figure}[htbp]
  \centerline{\plotone{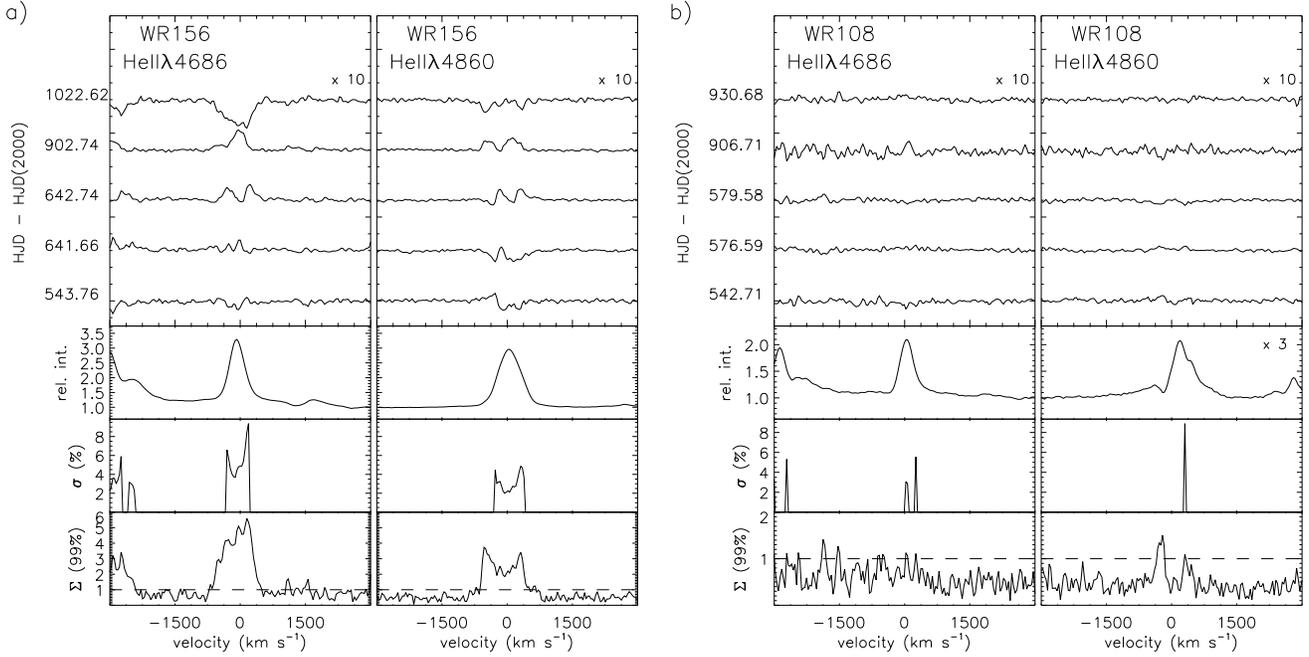}}
  \caption[Montage of line-profiles from a) WR~156 (WN8h) and b) WR~108 (WN9h)]{Same as Fig.\ref{fig1N} for a) WR~156 (WN8h) and b) WR~108 (WN9h).}
  \label{fig8N}
\end{figure}

\begin{figure}[htbp]
  \centerline{\plotone{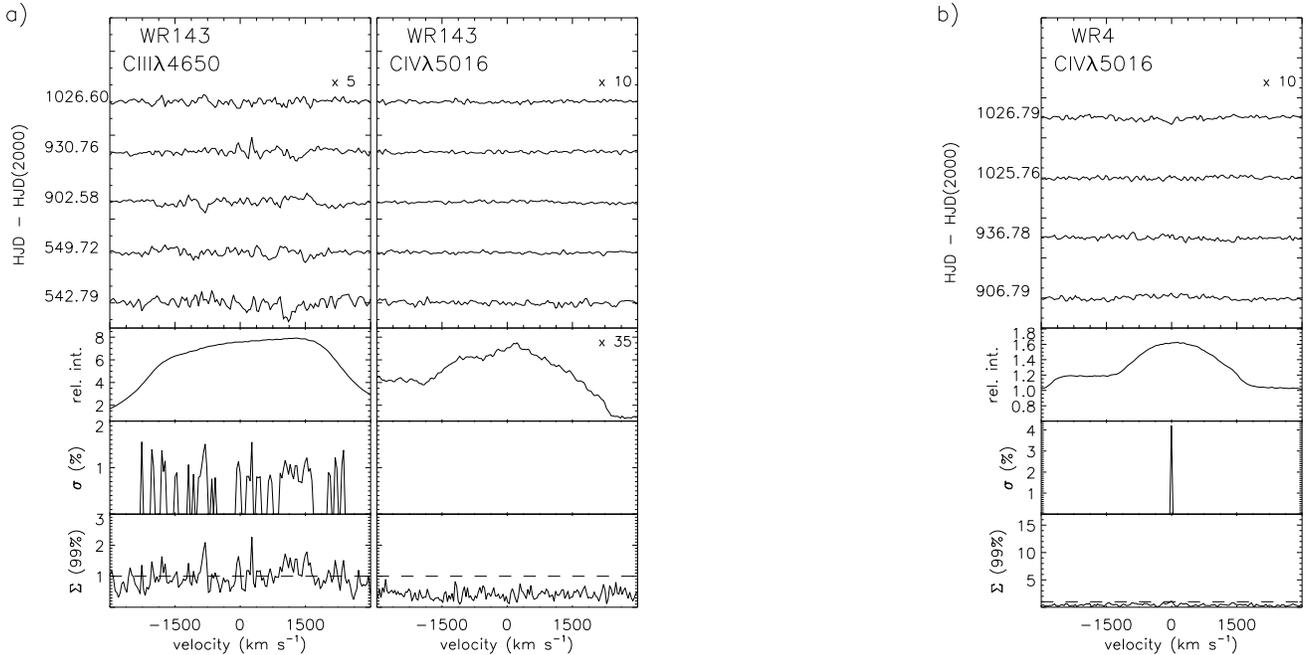}}
  \caption[Montage of line-profiles from a) WR~143 (WC4) and b) WR~4 (WC5)]{Same as Fig.\ref{fig1N} for a) WR~143 (WC4) and b) WR~4 (WC5) for the C{\sc iii}$\lambda$4650 (left) and C{\sc iv}$\lambda$5016 (right) line-profiles. Note that for WR~4, the montage for the C{\sc iii}$\lambda$4650 line is not shown as the data are unusable.}
  \label{fig9N}
\end{figure}

\begin{figure}[htbp]
  \centerline{\plotone{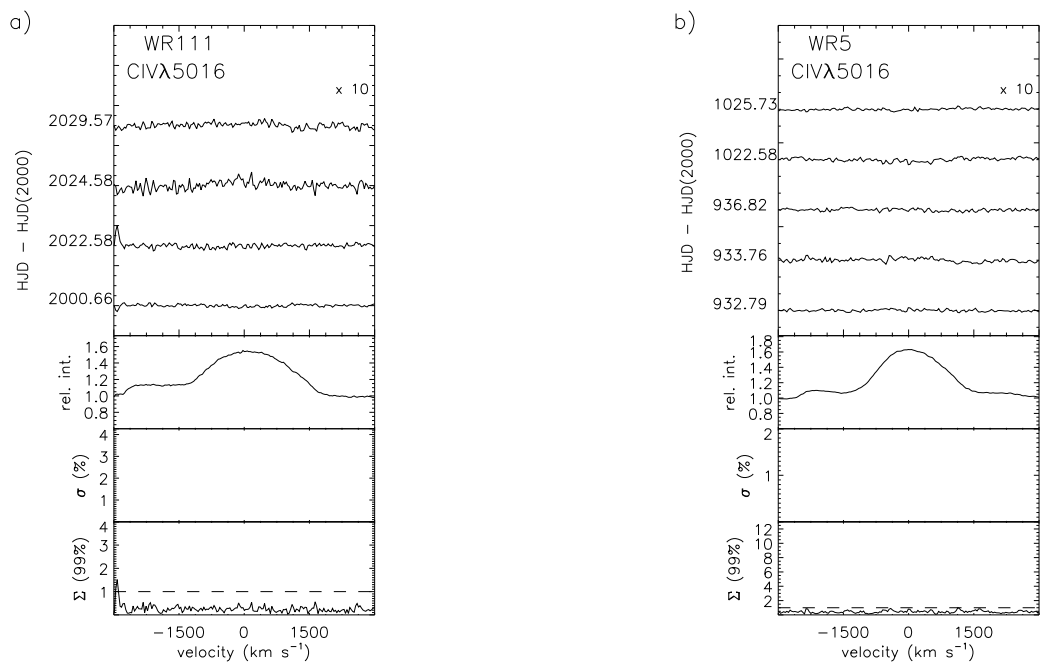}}
  \caption[Montage of line-profiles from a) WR~111 (WC5) and b) WR~5 (WC6)]{Same as Fig.\ref{fig9N} for a) WR~111 (WC5) and b) WR~5 (WC6). Note that the montage for the C{\sc iii}$\lambda$4650 line is not shown as the data are unusable.}
  \label{fig10N}
\end{figure}

\begin{figure}[htbp]
  \centerline{\plotone{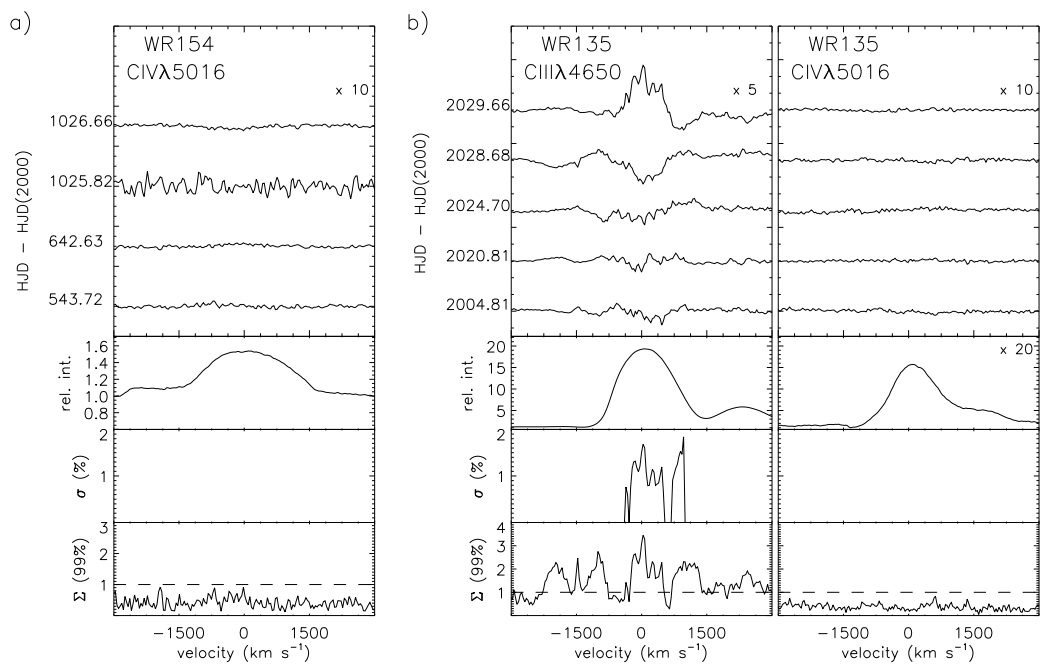}}
  \caption[Montage of line-profiles from a) WR~154 (WC6) and b) WR~135 (WC8)]{Same as Fig.\ref{fig9N} for a) WR~154 (WC6) and b) WR~135 (WC8). Note that for WR~154, the montage for the C{\sc iii}$\lambda$4650 line is not shown as the data are unusable.}
  \label{fig11N}
\end{figure}

\begin{figure}[htbp]
  \centerline{\plotone{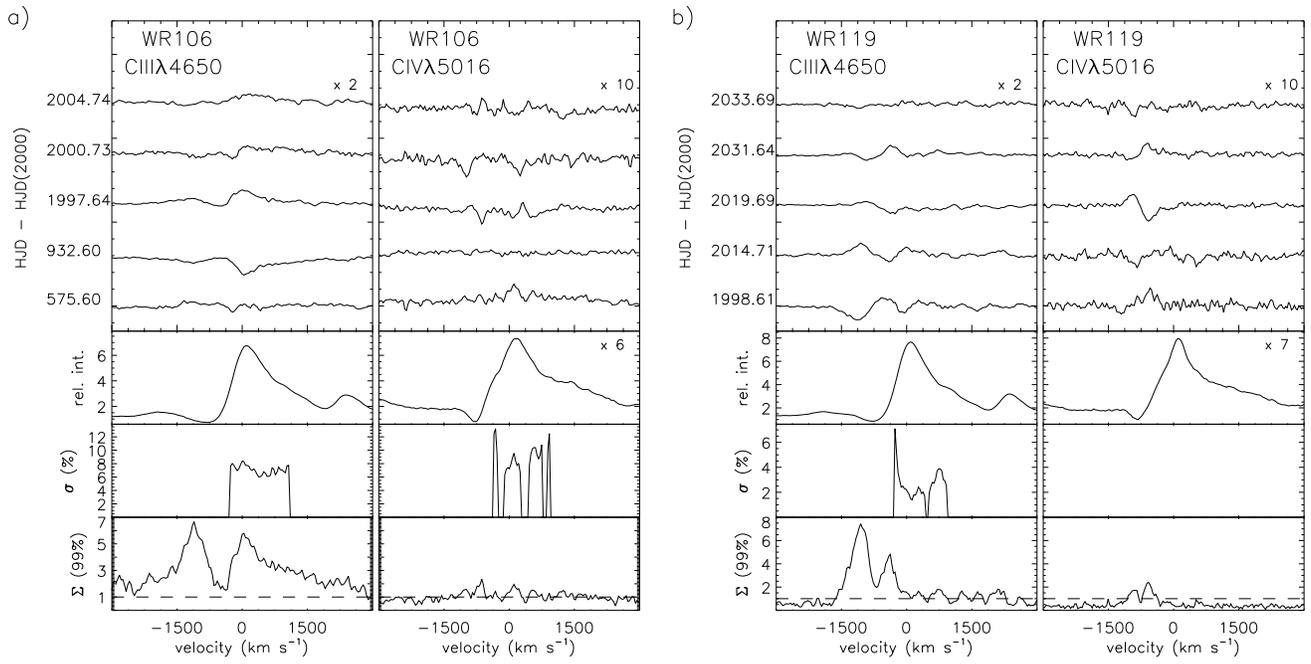}}
  \caption[Montage of line-profiles from a) WR~106 (WC9d) and b) WR~119 (WC9d)]{Same as Fig.\ref{fig9N} for a) WR~106 (WC9d) and b) WR~119 (WC9d).}
  \label{fig12N}
\end{figure}

\begin{figure}[htbp]
  \centerline{\plotone{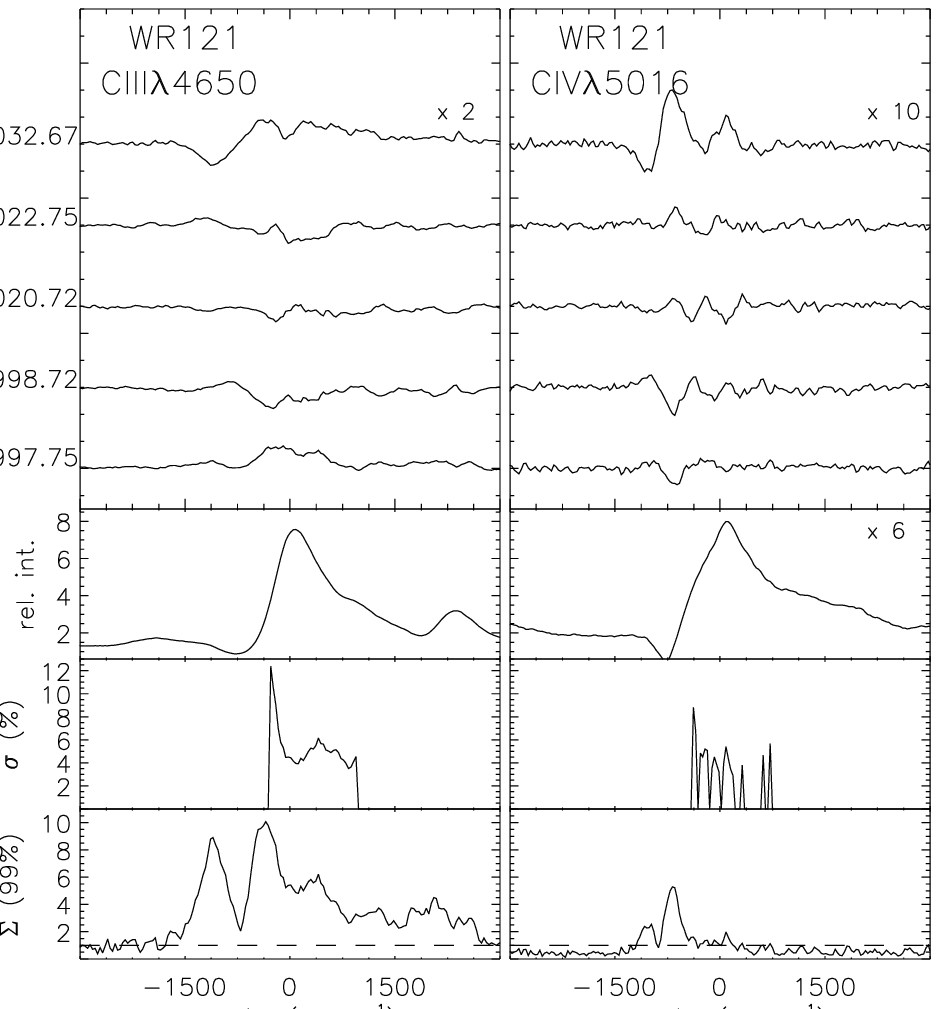}}
  \caption[Montage of line-profiles from WR~121 (WC9d)]{Same as Fig.\ref{fig9N} for WR~121 (WC9d).}
  \label{fig13N}
\end{figure}

The final spectra are shown in the top panels of Figures~\ref{fig1N}
to \ref{fig13N} for the two strongest lines present in our wavelength interval.  For WN stars these are 
He{\sc ii}$\lambda$4686 and He{\sc ii}$\lambda$4860 while for WC 
stars they are C{\sc iii}$\lambda$4650 and C{\sc iv}$\lambda$5016.  
For each star we plot the differences between individual 
spectra and the global mean (which is shown in the second panel from the 
top). The differences have been shifted vertically for clarity.  The Heliocentric 
Julian Date (HJD) of each observation is indicated on the left-hand side of
the plot and the scale factor of the ordinate is indicated in the top right-hand
corner of each panel. The graphs are organised by spectral type starting with WN stars 
from early to late types (Figures~\ref{fig1N} to \ref{fig8N})
followed by WC stars from early to late types (Figures~\ref{fig9N} to
\ref{fig13N}).  Note that for WR\,4, WR\,111, WR\,5 and WR\,154, we do not
show the plot of the C{\sc iii}$\lambda$4650 line because in those
cases the line was on the extreme edge of the wavelength interval and
therefore there was not enough continuum on one side of the line to
allow for an accurate enough spectrum rectification. This
could lead to a false detection of line variability.  We prefer to adopt
a conservative approach in this respect.

\clearpage

In order to search for significant spectroscopic variability among our
targets, we have adopted the formalism of \citet{Ful96}.  First, we
have calculated the temporal variance spectrum (TVS), which is a
statistically rigorous means of identifying the variable parts of our
spectra. For $N$ spectra, the TVS at wavelength $j$ is defined by:
\begin{eqnarray}
(TVS)_j={\frac{1}{(N-1)}}{\sum_{i=1}^N{\frac{w_i}{\alpha_{ij}}(S_{ij}-\bar{S}_j)^2}},
\end{eqnarray}
where the weights of the individual spectra, $w_i$, are given by
$w_i=(\frac{\sigma_0}{\sigma_{ic}})^2$ with
$${\sigma_{0}}^2=\left(\frac{1}{N}\sum^N_{i=1}{{\sigma_{ic}}^{-2}}\right)^{-1},$$ a
standardized dispersion. Here, $\bar{S}_j$ is the weighted mean
spectrum,  $\sigma_{ic}$ is the noise in the continuum  and
$\alpha_{ij}$ are wavelength correction factors given by
$\alpha_{ij}=(\frac{\sigma_{ij}}{\sigma_{ic}})^2$, $\sigma_{ij}$ being 
the noise associated with the flux $S_{ij}$.

Then, we have established the level of significant variability
assuming that our data are governed by a reduced chi-squared
distribution with N$-$1 degrees of freedom.  Using the standardized
dipersion, $\sigma_0$, for scaling, we therefore adopt a threshold of
: $\sigma_0^2 \chi^2_{N-1}(99\%).$ Finally, we used
$\Sigma_j(99\%)=\sqrt{\frac{(TVS)_j}{\sigma_0^2 \chi^2_{N-1}(99\%)}}$
to quantify the level of variability at each wavelength: a spectrum
that reaches a level of 1, is variable at a 1 $\sigma$ level, if it
reaches 3 it is variable at a 3 $\sigma$ level, etc. We have chosen a
confidence level of 99\%, which means that if the value of
$\Sigma_j(99\%)$ reaches 3, we are 99\%\ confident that the spectrum
is variable at a 3 $\sigma$ level. We note that as the standardized dispersion, $\sigma_0$ is
calculated using the noise level in the continuum regions and as the spectra are rectified, there is no real 
information on the variability of the continuum
regions. In the bottom panels of Figures~\ref{fig1N} to \ref{fig13N},
we have plotted the $\Sigma (99\%)$ spectrum for each star for easy
comparison with the spectra plotted above.

Inspection of Figures~\ref{fig1N} to \ref{fig13N} reveals that the majority of the 
stars in our sample show significant variability in at least one of the two main 
spectral lines in their spectrum. One notable exception is the WN2 star WR\,2. 
For WR\,4, WR\,5, WR\,111 and WR\,154, we did not detect
variability in the weak C{\sc iv} line, but we only have limited information
at our disposal since we do not have usable
data in the strong C{\sc iii} line. It is clear that the level of
variability is quite different from one star to another. The
variations range from a few $\sigma$s to close to 60$\sigma$s in
WR\,115. As the quality of our spectra is relatively uniform, we can,
in principle, compare the $\sigma$ from star to star, but its value
only provides information on the significance of the changes with
respect to the noise level of our data.

As the line strengths are very different from star to star, comparing
the significance level of the variations does not provide useful
physical information. A quantity that is potentially more interesting
is the fraction of the line flux that is variable. To characterize
this, it is best to use a quantity that is representative of the
variance in pixel j compared to the noise at that wavelength ($\sigma
_{ij}$) rather than use a weight that is linked to the noise in the
associated continuum, as the TVS does. To do this, \citet{Che07}
 has defined a modified TVS as follows:
\begin{eqnarray}
(TVS_{mod})_j=\frac{1}{(N-1)} \sum_{i=1}^N{\left(\frac{\sigma_{0j}}{\sigma_{ij}}(S_{ij}-\bar{S}_j)\right)^2},
\end{eqnarray}
where $\sigma_{0j}$ is the reciprocal of the $rms$ of the signal-to-noise 
ratio at pixel $j$ of the spectral line series.

In order to quantify what fraction of the line flux (without the
contiuum) is variable, we divide this value by the line flux,
$(\bar{S}_j -1)$.  Of course this quantity has no meaning outside
emission lines and in particular in continuum regions. To avoid any
problems, we therefore limited our caculations to a width corresponding to
a relative intensity level of 1.3 of each line.  Also, for very weak lines 
($<$ 1.3 in relative intensity), our ratio becomes meaningless because the
line flux, the denominator, is too close to zero and the variability level, the 
numerator, is dominated by the noise. In such cases, we have left the 
panel blank. Finally, absorption components of P~Cygni profiles are also excluded
since in that case the value of $(\bar{S}_j -1)$ will be negative and
cannot be compared directly with what is observed in emission
lines.

The values of $\sigma_j=(TVS_{mod})^{1/2}_j/(\bar{S}_j -1)$ are plotted
in the middle panels of Figures~\ref{fig1N} to \ref{fig13N}. Several
interesting conclusions can be drawn from these plots. First, for all
types of stars, both lines show roughly the same level of
variability. However, one has to note that in the case of WN stars, we
compared lines of the same ion and therefore it is likely that
in a given star, these two lines are formed in similar regions of the wind. The line widths 
are indeed, in most cases, very similar. Another interesting
feature of the variability is that the calculated values of $\Sigma$
are much stronger in P~Cygni absorption components than in the
corresponding emission components, i.e. the variations are more
significant. This is perhaps not surprising since that part of the
profile comes from a much smaller volume of the wind than the
emission and therefore suffers less from cancelling effects from
large-scale changes in density or ionisation arising in different
parts of the wind. This effect is clearly present in the P~Cygni
profile of WR\,123 (He{\sc ii}$\lambda$4686) or WR119 (C{\sc
iii}$\lambda$4650), for example.

After a careful examination of Figures \ref{fig1N} to \ref{fig13N}, we
divide our sample in three main categories: stars showing no profile
variability (NV), stars showing small-scale profile variability (SSV --- $\sigma < 5\%$) 
and stars showing large-scale variability (LSV --- $\sigma > 5\%$). Note that the stars
designated as NV are not necessarily stars that do not show spectral variability, 
but rather that no significant variability can be detected at a level of confidence of
99\% on spectra having signal-to-noise ratio $\sim$100. Our
conclusions are listed in the last column of Table~1. The detected subpeaks
have widths that range from $\sim$140 km\,s$^{-1}$ for the smallest isolated
structures to $\sim$350 km\,s$^{-1}$ for the largest.  Unfortunately, our limited 
dataset does not allow us to search for correlations between these widths
and, for example, the level of variability or the width of the global line.
In the following sections we will discuss in turn the variability we found in 
WN and WC stars.

\subsection{WN Stars} \label{WNs}

Significant variability has been found in all but one WN star in our
sample, i.e. WR\,2, the WN star of the earliest type. However, this
seems to be an unusual WR star. \citet{Ham06} note that this star has
unique line-profile shapes. These authors were unable to reproduce its
spectrum using their model atmospheres without convolving
the normal wind lines with a rotation profile with a velocity of 1900
km s$^{-1}$, which is very close to the star's critical velocity.  This is, of course, a 
simplistic way of including the effects of rotation on the line profile and is
likely to lead to improbable conclusions but it reflects the fact that the line-formation
process for this star is not totally understood and that the star is unusual
in that respect.

For WR\,3, WR\,152, WR\,128, WR\,136, WR\,131, WR\,158 and WR\,108,
variability has indeed been detected but at a relatively low level of
$\sigma\sim 2-3\%$ of the line flux. The nature of the changes is
also very different from the CIR-type variability we first set-out to identify.  
The structures are small and are most likely stochastic changes associated with
inhomogeneities in the wind of WR stars, such as
discussed from an observational point of view by \citet{Mof88} 
and \cite{Lep96}, for example.   More recently, there have been some 
insightful theoretical investigations into small-scale inhomogeneities in the winds of hot stars. \citet{des02a,des02b,des03,des05a,des05b} present radiation hydrodynamic simulations
of hot-star winds line-profile variations that arise as a consequence of the intrinsically unstable nature of the line-driving 
mechanism of the wind. Those calculations show, for example, that including Rayleigh-Taylor
or thin-shell instablities together with some lateral averaging of the diffuse radiation
is important to understand the behaviour of the width of the subpeaks as a
function of their position on the line profile, i.e. the lateral versus radial size
of the small scale inhomogeneities in the wind.  We believe that the characteristics
of the variations seen in the stars that we classify as SSV, are compatible with 
the type of changes caused by such small-scale inhomegeneities in the wind.

For WR\,1, WR\,110, WR\,115, WR\,134, WR\,120, WR\,123, WR\,124 and
WR\,156, the variations have a much larger amplitude. Therefore, it is
quite likely that the physical phenomenon from which they originate is
different from what is observed for SSV stars. We discuss these stars
in more detail below.

WR\,134 is well-known to show CIR-type variability for which a period of 2.3
days has been identified \citep[e.g.][]{McC94, Mor99}. We use it as a
test that indeed with 4--5 spectra we can identify easily the type of
variability we are looking for. One can clearly see on top of both the
He{\sc ii}$\lambda$4686 and He{\sc ii}$\lambda$4860 lines, broad emission
excesses at different positions on the line. The changes are highly
significant and reach $\sigma$~=~6--8 \%\ of the line flux.

Three other WN stars show a similar type of variability: WR\,1,
WR\,115 and WR\,120. These stars clearly show CIR-type variations; large-scale
structures are seen superposed on the broad emission lines reaching
$\sigma\sim$7--8\%\ of the line flux in WR\,1 and WR\,120 and
$\sigma\sim$10$-$12\%\ of the line flux in WR\,115. We therefore
classify these stars as LSV and further put them in the CIR-type
category.

The situation is less clear for WR\,110. Clear structures can be
distinguished on both spectral lines. The variations reach a level of
$\sigma\sim$5\%\ of the line flux, which is significantly higher than
the other stars we have classified as SSV and that we associate with
small-scale wind inhomogeneities. However, the detected
structures are not quite as strong as for those showing variability thought to 
be associated with CIRs. Therefore, we classify this star as a SSV and use it to
establish our approximate upper limit for the level of
small-scale/blob-like variability.

Finally, for WR\,123, WR\,124 and WR\,156, all WN8-type stars, a
different type of variability is found. The changes reach a relatively
high level (from 5 to 10\% of the line flux) but appear to be 
different in nature from those thought to be associated with CIRs. The $\sigma$ profile
has a very typical shape with two maxima, one on the red and the other
on the blue side of the emission line and a minimum at line center.
The only exception is the He{\sc ii}$\lambda$4686 profile of WR123 but
in that case there is a very strong contrast between the variability
in the absorption component and that of the emission. In fact this
effect of a more variable absorption component seems to be amplified
in WR\,108, the only WN9 star in our sample.  Indeed, despite the fact
that we have classified this star as SSV, the changes are different
in nature from that observed in the other stars we have put in that
category. In this case, the variability is actually confined to the
P~Cygni absorption component of the He{\sc ii}$\lambda$4860 line,
which is clearly present. In view of the specific nature of the
changes, we create a separate class of LSV for WN8-type stars.
Actually, WN8 stars are well-known to display an unusual
behaviour. \citet{Mar98} presented a comprehensive study of the
photometric and spectroscopic variability patterns of 10 Galactic WN8
stars and found a link between the level of variability in the
continuum and in spectral lines which they suggest may be linked to
pulsational instabilities. Other unusual behaviours of WN8 stars as
summarized by \citet{Mar98} include a low binary frequency, a high
distance from the Galactic plane and/or runaway speeds and the fact
that they are rarely found in stellar clusters and
associations. Therefore, it is perhaps not surprising to find that
they fall in a class of their own compared with other types of WR
stars when it comes to spectral variability.

\subsection{WC Stars} \label{WCs}

Because of problems with our spectrum rectification procedure of the
C{\sc iii}$\lambda$4650 lines in our early-type WC stars, the
information at our disposal for these stars is more limited than for
WN stars. Indeed, this line is located very close to the blue
extremity of our wavelength interval.  Therefore, the continuum
regions that are available bluewards of the line are much more
restricted. This reduces the accuracy that can be achieved during the
spectrum rectification procedure. Of course, the higher the contrast
of the line with respect to the continuum, the more it is affected by
this effect. The WC4 star WR\,143 has a somewhat weaker C{\sc iii} line
than the other early WC stars and therefore we were able to obtain a
sufficiently accurate rectified line profile. Note that this line is slightly 
affected by blending with the neighbouring He{\sc ii}$\lambda$4686 line.
This is the reason why the line rises towards the red.  However, the C{\sc
iv}$\lambda$5016 line in that star is practically inexistent. We
classify this star as SSV as its variability level is $\sim 1\%$ of
the line flux.

For WR\,4, WR\,111, WR\,5 and WR\,154, all WC5 and WC6 stars, no
variability is detected in the weak C{\sc iv}$\lambda$5016 line, which is
the only usable line in our wavelength range. We therefore classify
them as stars showing no profile variabilities (NV) but note that new 
observations of a stronger line should be obtained to verify this assertion.

For WR\,135, we find small-scale variability only in the C{\sc
iii}$\lambda$4650 line. The C{\sc iv}$\lambda$5016 line is constant at
the level of accuracy of our dataset. This star has been clearly
identified as a typical, small-scale variable by \citet{Rob92} and
\citet{Lep96}.  We reach the same conclusion here; we classify it as a
SSV.

Finally, the three WC9d stars, WR\,106, WR\,119 and WR\,121, all reach a
high level of variability. However, as indicated by their spectral
types, these WC stars are thought to produce carbon-based dust in
their atmosphere. Some WC dust makers have been shown to be binaries
\citep[e.g. the well-known system WR\,140;][]{Hac,Wil}. In that case,
it is believed that the dust is produced within the cone-shaped shock
formed as a consequence of the collision between the two stellar
winds.  As the gas flows along the shock cone, it eventually reaches 
distances where the density compression is still adequate for dust 
formation and the intensity of the ultraviolet flux from the star is sufficiently 
low not to destroy it. For single stars,
however, a definite mechanism to form dust in the wind has yet not
been identified. Shocks within the wind itself caused by radiative
instabilities have been claimed to be possible sources
\citep{Zub1,Zub2}.  Since all three of our WC9d stars are variable with a large amplitude, we
believe that caution must be used, since colliding-wind binaries have
been shown to display line-profile variability in the optical with 
characteristics very similar to that of CIRs \citep[e.g.][]{Hil00}. It is therefore 
possible that the WC9d stars in our sample are yet un-identified binaries. We plan to
investigate this further by carrying out intensive spectroscopic and
photometric monitoring campaigns of these stars. Only a detailed study can reveal
if the changes are due to a colliding-wind shock-cone or to CIRs. Therefore, we
classify these stars as LSV but create a separate class, the WC-late
dust makers.

\section{Discussion \&\ Conclusion}

The results of our search for optical spectroscopic variability among presumably single WR stars 
can be summarised as follows. For eight stars, WR\,3, WR\,152, WR\,128, 
WR\,136, WR\,131, WR\,158, WR\,108 and WR\,135, variability was found but at a small 
level compared to the flux of the line. This type of variability is most-likely associated with inhomogeneities 
in the dense WR wind, which move with it and reveal themselves as small-scale structures 
superposed on the broad emission lines.  In more detailed studies, they have been found to
move from line center to the edges of the lines as described for example by \citet{Mof88} and 
\cite{Lep96}. As the driving mechanism for these radiatively-driven winds has been found to
be inherently unstable \citep{Owo88}, such variability is expected for all WR stars.

Nevertheless, for 5 other stars in our sample, namely WR\,2, WR\,4, WR\,111, WR\,5,
WR\,154 we found no significant line-profile variability. Note however that in the last 4
cases, our conclusion is based solely on a single relatively weak
line. In fact, WR\,111, is a star that is known to display small-scale spectral variability due to the 
presence of inhomogeneities in its wind and has been studied by \cite{Lep99}.
In any case, identifying a star as non-variable always depends on the
quality of the data in hand. Down to what level of accuracy these stars are constant
is, indeed, an important question. One would like to verify if {\it all} WR stars
have inhomogeneities in their winds, if some stars present a different level
of variability and if so, what is the cause of this difference. This kind of information can 
help us better understand the physical mechanisms
responsible for these unstable outflows.

Nine stars in our sample presented large-scale profile variability. Of these,
we have identified three new candidates for CIR-type variability:
WR\,1, WR\,115 and WR\,120. Follow-up observations with a high temporal
resolution will be required to determine if these changes are periodic
and if so identify the period. We have done so already for WR\,1 and present our
results in an upcoming paper \citep{CheST09b}. 

Although the results presented here are not yet for our complete sample, it is interesting to 
compare our findings to what is known for OB stars. Although the presence of DACs in O-star P Cygni 
profiles is found to be common \citep[e.g.][]{How89}, cyclical variability in 
emission lines is not found to show the same ubiquitous behavior. \cite{Mor04}
present a long-term monitoring campaign of the H$\alpha$ line of 22 bright OB supergiants.
Variability was found in all cases with two possible origins: either the underlying photospheric
profile and/or a wind component. Cyclical changes which very likely originate from large-scale
structures in the wind were found for only two stars in their sample ($\sim$ 14\%\ of the stars).  In
this study we have identified 4 CIR-type variables (3 of which are new) out of a sample of 25, or $\sim$ 16\% ,
a very similar fraction. Alternatively, we can exclude the WN\,~8 and WC\,~9 stars from this estimate;
if we do so we also obtain a similar fraction, 4/19$\sim$21\%. This striking difference between variability in P~Cygni  
absorption components and emission lines is most likely due to the fact in the former we
are sampling a much more physically restricted volume of the wind (i.e. that in front of the
star) compared to the entire wind which can be subject to cancelling effects in parts of the 
wind which are physically distinct but that are travelling at a similar projected velocity.

In addition to the stars showing variability thougt to be associated with CIRs, we have 
defined two distinct classes of LSV: WN8 stars which are already known to be a peculiar
class of WR stars and WC9d stars. In the latter case, it
remains to be checked if the changes originate in the wind of a single star
because of the presence of CIRs or if these stars are in fact colliding-wind 
binaries. If of binary origin, a very similar variability pattern is expected 
with some differences that can be identified with a detailed monitoring campaign.
For example, strong epoch-dependancy in spite of the presence of periodic changes would
suggest a single-star origin. Also, shock-cones translate into a very distinct and easily
recognisable variability pattern on spectral lines, translating to at most, two large excess emissions
moving around periodically on the emission line.

Although our survey has revealed a relatively small fraction ($\sim$12\%) of stars showing
line-profile variability thought to be associated with CIRs among WR stars, it still provides 
some new candidates for potential estimates of WR rotation velocities.  It also provides us 
with the opportunity to study in more detail the  CIR phenomenon for which still very little is 
known by providing a larger pool of candidates to observe.  It is clear that better 
observational constraints of this phenomenon are crucial to reach a greater understanding 
of the physical mecanisms from which they originate.

\acknowledgments

We wish to thank G. Skalkowsky, C. Foellmi and G. Caron for participating in observing
campaigns in 2001/2002. We are also grateful to the (anonymous) referee for comments
that led to an improvement of the quality of this paper. Finally, we would like to thank the
National Sciences and Engineering Research Council (NSERC) of Canada
for financial support.


\clearpage

\end{document}